\documentclass[aps,prl,showpacs,amsmath,amssymb,amsfonts,twocolumn,
superscriptaddress,floatfix,footinbib]{revtex4}

\usepackage{subfigure}
\usepackage{graphicx}
\usepackage[flushmargin]{footmisc}
\usepackage{color}
\usepackage{soul}
\usepackage{longtable}
\usepackage[normalem]{ulem}

\begin{document}

\title{Polariton Condensation in photonic molecules}

\author{Marta Galbiati}
\affiliation{CNRS-Laboratoire de Photonique et Nanostructures, Route
de Nozay, 91460 Marcoussis, France}
\author{Lydie Ferrier}
\affiliation{CNRS-Laboratoire de Photonique et Nanostructures, Route
de Nozay, 91460 Marcoussis, France}
\author{Dmitry D. Solnyshkov}
\affiliation{LASMEA, CNRS, Clermont University, University Blaise Pascal, 24 avenue des
Landais, 63177 Aubi\`{e}re cedex, France}
\author{Dimitrii Tanese}
\affiliation{CNRS-Laboratoire de Photonique et Nanostructures, Route
de Nozay, 91460 Marcoussis, France}
\author{Esther Wertz}
\affiliation{CNRS-Laboratoire de Photonique et Nanostructures, Route
de Nozay, 91460 Marcoussis, France}
\author{Alberto Amo}
\affiliation{CNRS-Laboratoire de Photonique et Nanostructures, Route
de Nozay, 91460 Marcoussis, France}
\author{Pascale Senellart}
\affiliation{CNRS-Laboratoire de Photonique et Nanostructures, Route
de Nozay, 91460 Marcoussis, France}
\author{Isabelle Sagnes}
\affiliation{CNRS-Laboratoire de Photonique et Nanostructures, Route
de Nozay, 91460 Marcoussis, France}
\author{Aristide Lema\^{i}tre}
\affiliation{CNRS-Laboratoire de Photonique et Nanostructures, Route
de Nozay, 91460 Marcoussis, France}
\author{Elisabeth Galopin}
\affiliation{CNRS-Laboratoire de Photonique et Nanostructures, Route
de Nozay, 91460 Marcoussis, France}
\author{Guillaume Malpuech}
\affiliation{LASMEA, CNRS, Clermont University, University Blaise Pascal, 24 avenue des
Landais, 63177 Aubi\`{e}re cedex, France}
\author{Jacqueline Bloch}
\affiliation{CNRS-Laboratoire de Photonique et Nanostructures, Route
de Nozay, 91460 Marcoussis,
France}\email[]{jacqueline.bloch@lpn.cnrs.fr}

\date{\today}

\begin{abstract}
We report on polariton condensation in photonic molecules formed by
two coupled micropillars. We show that the condensation process is
strongly affected by the interaction with the cloud of uncondensed
excitons. Depending on the spatial position of these excitons within
the molecule, condensation can be triggered on both binding and
anti-binding polariton states of the molecule, on a metastable state
or a total transfer of the condensate into one of the micropillars
can be obtained. Our results highlight the crucial role played by
relaxation kinetics in the condensation process.
\end{abstract}

\pacs{71.36.+c, 67.85.Hj, 78.67.Pt, 78.55.Cr}
%\keywords{}

\maketitle

Most of the experimental studies in atomic Bose condensates have
explored conditions of thermodynamic equilibrium since typical
condensate lifetimes are much longer than interaction times. Recent
theoretical proposals have shown that out of equilibrium bosonic
systems present qualitatively new behaviors~\cite{Werner2005}. One
proposed way to reach this regime is the use of photonic systems
with effective photon-photon interactions and dissipation provided
by inherent optical losses~\cite{Gerace}. Localized to delocalized
phase transitions~\cite{Greentree2006,Schmidt2010}, highly entangled
states~\cite{Hartmann2006}, or fermionisation effects in a ring of
coupled sites~\cite{Carusotto2009} are predicted in such systems.

Microcavity polaritons are a model system for the investigation of
the physics of driven-dissipative boson
condensates~\cite{Kasprzak2006,Balili2007,Lai2007,Christopoulos2007,
BajoniPRL2008,Wertz2010,Cerda2010}. They are the quasi-particules
arising from the strong coupling between
 excitons confined in quantum wells and the
optical mode of a microcavity. Because of their light-matter nature,
polaritons present peculiar properties: they interact efficiently
with their environment through their excitonic
part~\cite{Tassonne-polaritonpolariton,Ciuti-parametric} while their
photonic part enables efficient coupling with the free space optical
modes. Polariton condensates can be generated in zero dimensional
 micropillars~\cite{BajoniPRL2008} or in arrays of
pillars with fully controlled
coupling~\cite{Bayer1998,Guttroff2001}. In this configuration,
 the non-equilibrium nature of polariton
condensates should allow the realization of metastable collective
states, such as the self-trapped states in a bosonic Josephson
junction \cite{Smerzi, Sarchi2010, Shelikh2008}.

In the present paper we investigate polariton condensation in
photonic molecules obtained by coupling two micropillars. We
demonstrate that polariton interactions strongly affect the way
condensation occurs in such coupled system, not only modifying the
wavefunction of the polariton condensate, but also the relaxation
dynamics. This effect, specific to an out-of-equilibrium bosonic
system, is illustrated by considering different positions of the non
resonant excitation within the molecule. When the excitation spot is
placed at the center of the molecule, polariton condensation
 is observed on both binding
 and anti-binding states. Interactions induce strong changes
 in the condensate
  wavefunction, the most important one being the change in its spatial anisotropy.

  When the excitation spot is positioned on one of the two coupled
micropillars, condensation occurs in a very different way. As the
excitation power is increased, the polariton condensate is first
created in a metastable state localized in the excited pillar by an
effect similar to self trapping\cite{Smerzi}. Further increasing the
pumping power, the condensation tends to progressively evolve from a
kinetic regime to a regime closer to thermodynamical equilibrium:
massive occupation is observed on the lowest energy state of the
system, mainly localized in the non-excited pillar.
 These results are simulated using a relaxation model
 including semi-classical Boltzmann equations and nonlinear Schr\"{o}dinger equation in a self-consistent
 way. The key features of our experiments can only be reproduced if one properly includes the
  changes of the relaxation rates
 due to interaction-induced modifications of the overlaps between the different polariton states.
%These results highlight the key role played by the
%kinetics.

\begin{figure}[t]
\includegraphics[width= 0.8\columnwidth]{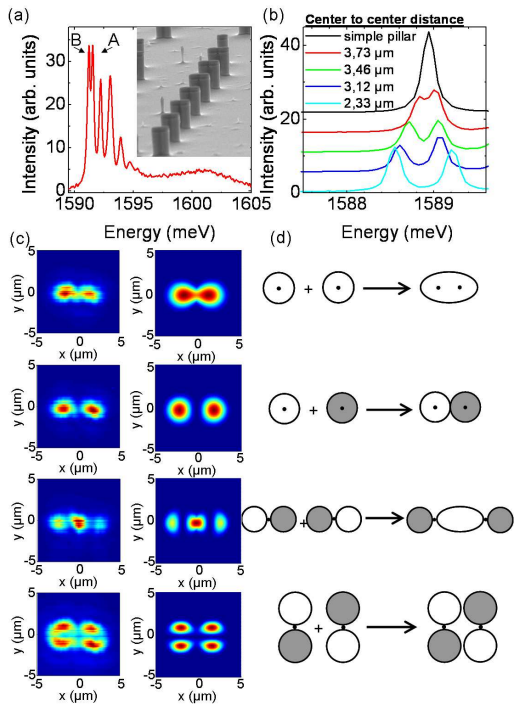}
\caption{(Color online) (a) Emission spectrum measured on a single
molecule at low excitation power ($d\,=\,4\,\mu m$ and
$d_{CC}\,=\,3.73\,\mu m$); A and B indicate the binding and
anti-binding states.
 Inset : scanning electron micrograph of an array of pillars and molecules. b)
  Emission spectra measured on a $4\,\mu\,m$ round micropillar (black line) and on photonic
  molecules with $d\,=\,4\,\mu m$ and various values of $d_{CC}$. c) left (resp. right) column :
measured (resp. calculated) emission pattern of the four lowest
energy modes of a photonic molecule with $d\,=\,4\,\mu m$ and
$d_{CC}\,=\,3.73\,\mu m$; d) Schematic of the hybridization of the
individual pillar modes within a photonic molecule.}\label{Fig1}
\end{figure}

  The microcavity sample, described in
Ref.~\cite{BajoniPRL2008}, consists in a $\lambda/2$ cavity with a
 quality factor exceeding $16\, 000$ and containing 12 GaAs quantum
wells. Coupled micropillars were fabricated using electron beam
lithography and Inductively Coupled Plasma dry etching. The diameter
$d$ of the micropillars ranges from $3$ to $4$ $\mu m$ and their
center to center distance $d_{CC}$ is varied from 2.3 to $3.7$ $\mu
m$, corresponding to an expected coupling constant between $0.1$ and
$1~meV$.  Note that $d_{CC}$ is always kept smaller than twice the
radius of the pillars, ensuring the direct coupling of the polariton
modes of the two micropillars. A scanning electron micrography of an
array of such photonic molecules is shown in Fig.\ref{Fig1}.a.
Microphotoluminescence experiments are performed on single molecules
using a single mode cw Ti:Saph laser focused onto a 2~$\mu$m
diameter spot with a microscope objective. The sample is maintained
at 10~K and the excitation laser energy is tuned typically 100~meV
above the lower polariton resonance, thus providing non-resonant
optical excitation of the system. The emission is collected through
the same objective and imaged on the entrance slit of a
monochromator. The spectrally dispersed emission is detected with a
nitrogen-cooled CCD camera. We define the detuning
$\delta=E_{c}-E_{x}$ as the energy difference between the lowest
energy photonic mode and the exciton resonance.

\begin{figure}[t]
\includegraphics[width= \columnwidth]{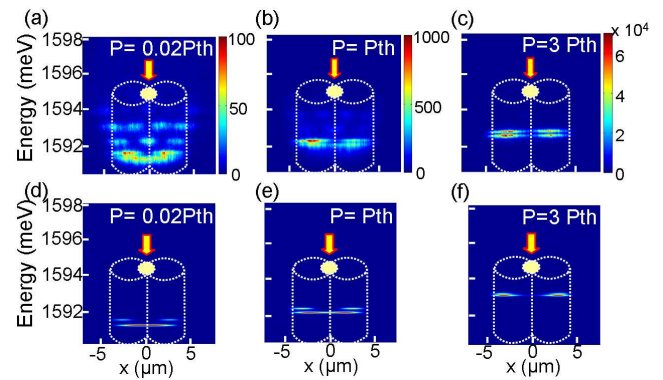}
\caption{(Color online) \textbf{Excitation at the center of the
molecule}: a)-c)Spectrally resolved emission distribution along the
molecule axis measured for several excitation powers; d)-f)
Calculated emission patterns for several excitation powers.
$d\,=\,4\,\mu m$ and $d_{CC}\,=\,3.46\,\mu m$. $\delta\,=\,-3\,
meV$}\label{Fig2}
\end{figure}

The polariton modes in these photonic molecules are investigated by
photoluminescence measurements at low excitation power. An example
of an emission spectrum measured on a single molecule is presented
in Fig.~\ref{Fig1}.a. Discrete emission peaks are observed
corresponding to polariton quantum states fully confined in the
microstructure. The broader line at higher energy is due to emission
of the excitonic reservoir. The two lowest energy modes (labeled B
and A) are attributed to the binding and anti-binding states arising
from the hybridization of the lowest energy mode of each
micropillar. The splitting between these states is proportional to
the coupling between the two micropillars. It can be continuously
tuned by changing the center to center distance $d_{CC}$, as
illustrated in Fig.~\ref{Fig1}.b. The left column of
Fig.~\ref{Fig1}.c shows the spatial distribution (near field) of the
four lowest energy modes in a molecule made of two $4 \,\mu m$
pillars with $d_{CC}= 3.73 \,\mu m$ (the first two panels correspond
to the B and A modes). The right column shows the calculated
polariton wavefunctions considering a confinement potential taken as
infinite outside the photonic molecule and equal to zero inside. As
schematically illustrated in Fig.~\ref{Fig1}.d, all these states
result from the hybridization of the optical modes of each
individual micropillar, showing the strong analogy between our
system and the orbitals of a diatomic molecule.

We now discuss polariton condensation in these molecules under non
resonant optical excitation. As we have shown in our recent work on
single micropillars~\cite{Ferrier2011}, such an excitation scheme
not only populates the confined polariton states,
 but also creates a population of
uncondensed excitons called the excitonic reservoir. Because of the
limited diffusion length of excitons, this reservoir remains
localized in the excitation area. Repulsive interactions between
polaritons and the excitonic reservoir strongly influence the
precise quantum state in which polaritons accumulate. In the
following we will consider two different locations of the excitonic
reservoir within the molecule, which can be selected via the
position of the excitation laser spot. First we will consider
excitation
 conditions where the reservoir of uncondensed excitons is at the center of the molecule. Then we will
 address the case of asymmetric excitation, in which the excitonic reservoir is injected only in one of the
 two micropillars. Specific spatial
behavior of the polariton condensates is observed in each case,
driven by the polariton interaction
 with the excitonic reservoir.

\begin{figure}[t]
\includegraphics[width= 0.8\columnwidth]{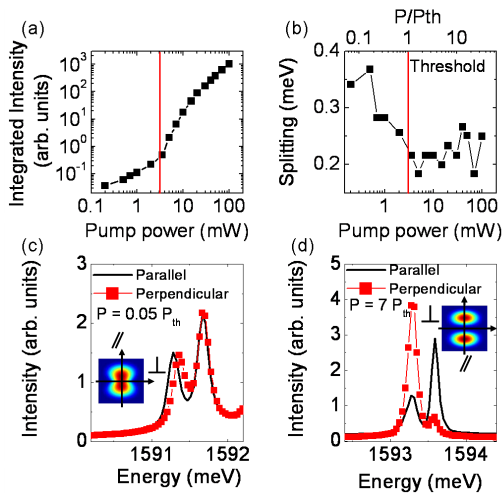}
\caption{(Color online) \textbf{Excitation at the center of the
molecule}: a) Integrated intensity measured as a function of the
excitation power; b) Splitting between the two lowest energy
emission lines measured as a function of the excitation power; c)
Lowest energy emission lines measured with a polarization parallel
(solid line) and perpendicular (squares) to the molecule axis for
$P\,=\,0.05\,P_{th}$; d) same as c) for $P\,=\,7\,P_{th}$. The inset
shows the calculated spatial shape of the binding state}\label{Fig3}
\end{figure}

Figure~\ref{Fig2}.a-c present the measured emission distribution
along the axis of a photonic molecule excited at its center. As
summarized in Fig.~\ref{Fig3}.a, above a well defined excitation
threshold, a strong non-linear increase of the emission intensity is
observed, one of the signatures of polariton
condensation~\cite{BajoniPRL2008}. Under this excitation condition,
a massive accumulation of polaritons occurs both in the binding and
anti-binding lowest energy polariton states. Interestingly, as the
excitation power is increased, a progressive change of the spatial
shape of the polariton wavefunction is observed, with vanishing
probability density at the center of the molecule. This behavior is
due to repulsive interactions with the excitonic reservoir injected
in that region.

To describe our experiments, we simulate polariton relaxation using coupled semi-classical Boltzman equations
 and nonlinear Schr\"{o}dinger equations in
 a self-consistent way~\cite{supplementary}. As described in the supplementary information,
 on one hand, a nonlinear Schr\"{o}dinger equation
 is solved at each time iteration, in the presence of the
 potential induced by the excitonic reservoir and by the occupation of the polariton
 states,
 to find the eigenstates of the system.
 On the other hand, the time evolution of these occupation
numbers are calculated using the semi-classical Boltzmann equations
in which the scattering rates are obtained using the exact shape of
the polariton states obtained from the non-linear Schr\"{o}dinger
equation. Both exciton-phonon and exciton-exciton scattering
mechanisms are considered. The reservoir is assumed to be
thermalized at the lattice temperature. The key ingredient to
describe our experiments is to consider scattering rates
proportional to the spatial overlap between initial and final
states. For instance, the scattering rate of two excitons from the
reservoir resulting in one polariton in state $i$ and one exciton in
the reservoir is proportional to
$\int|\psi_{i}(x,y)|^{2}(\rho_{R}(x,y))^{3}dxdy$, where
$\psi_{i}(x,y)$ is the wavefunction of the polariton state $i$ and
$\rho_{R}(x,y)$ is the spatial distribution of the excitonic
reservoir, described by a gaussian of width $w_R$. To fit the data
we use an exciton-exciton scattering rate $W_{XX}=2~10^{3}\,s^{-1}$
and an exciton-phonon scattering rate $W_{XP}= 10^9~$s$^{-1}$. The
lifetime of the reservoir is equal to $400\, ps$, and that of the
polariton states to $30\, ps$, a value extracted from coherence
measurements in ref.[\onlinecite{Wertz2010}].

The results of the model for central pumping are shown in
Fig.~\ref{Fig2}d-f. The potential of the reservoir increases with
the optical pumping and has a maximum at the center of the molecule,
inducing the spatial separation of the states of each micropillar.
As a result, both the coupling and the associated splitting between
the two lowest energy states decrease. Indeed, this feature is shown
in Fig.~\ref{Fig3}.b: simultaneous to the spatial separation of the
polariton distributions, the measured splitting goes down to a
minimum at about $P=4 \,P_{th}$. At higher powers, the measured
splitting tends to increase again. In the following, we will show
that at such high density the observed splitting is no longer
related to the binding anti-binding splitting but is an anisotropic
splitting induced by the lateral spatial shrinking of the condensate
wavefunction in each micropillar. This feature is not reproduced by
our model which does not include the polarization degree of freedom.

Polarization resolved measurements were performed for different
excitation densities. Figure~\ref{Fig3}.c and \ref{Fig3}.d display
emission spectra linearly polarized parallel and perpendicular to
the molecule axis, below and above threshold. In the low density
regime (Fig.~\ref{Fig3}.c), the presence of uncondensed excitons has
a negligible effect, and we observe a polarization splitting in both
the binding and anti-binding states. As previously reported
theoretically and experimentally~\cite{Vasconcellos2011}, this
splitting is larger for the binding state and amounts to $70\, \mu
eV$. The inset of Fig.\ref{Fig3}.c shows that the overall shape of
the polariton binding state is elongated along the molecule. This is
why the lowest energy line of the binding state doublet is polarized
parallel to the molecule axis.
 The situation is different at high
density ($P>4 \, P_{th}$), where a stronger potential barrier is
induced by the excitonic reservoir at the center of the molecule. In
this case, the coupling between the states of each micropillar
becomes too small to be resolved in our experiment. The remaining
splitting is then purely a polarization splitting. In each
micropillar, the polariton wavefunction is shrunk by the excitonic
repulsive potential and thus it is more elongated perpendicularly to
the molecule axis. This change in anisotropy is evidenced in the
polarization of the emission.The lowest energy emission line is now
strongly polarized perpendicular to the molecule axis.

\begin{figure}[t]
\includegraphics[width= \columnwidth]{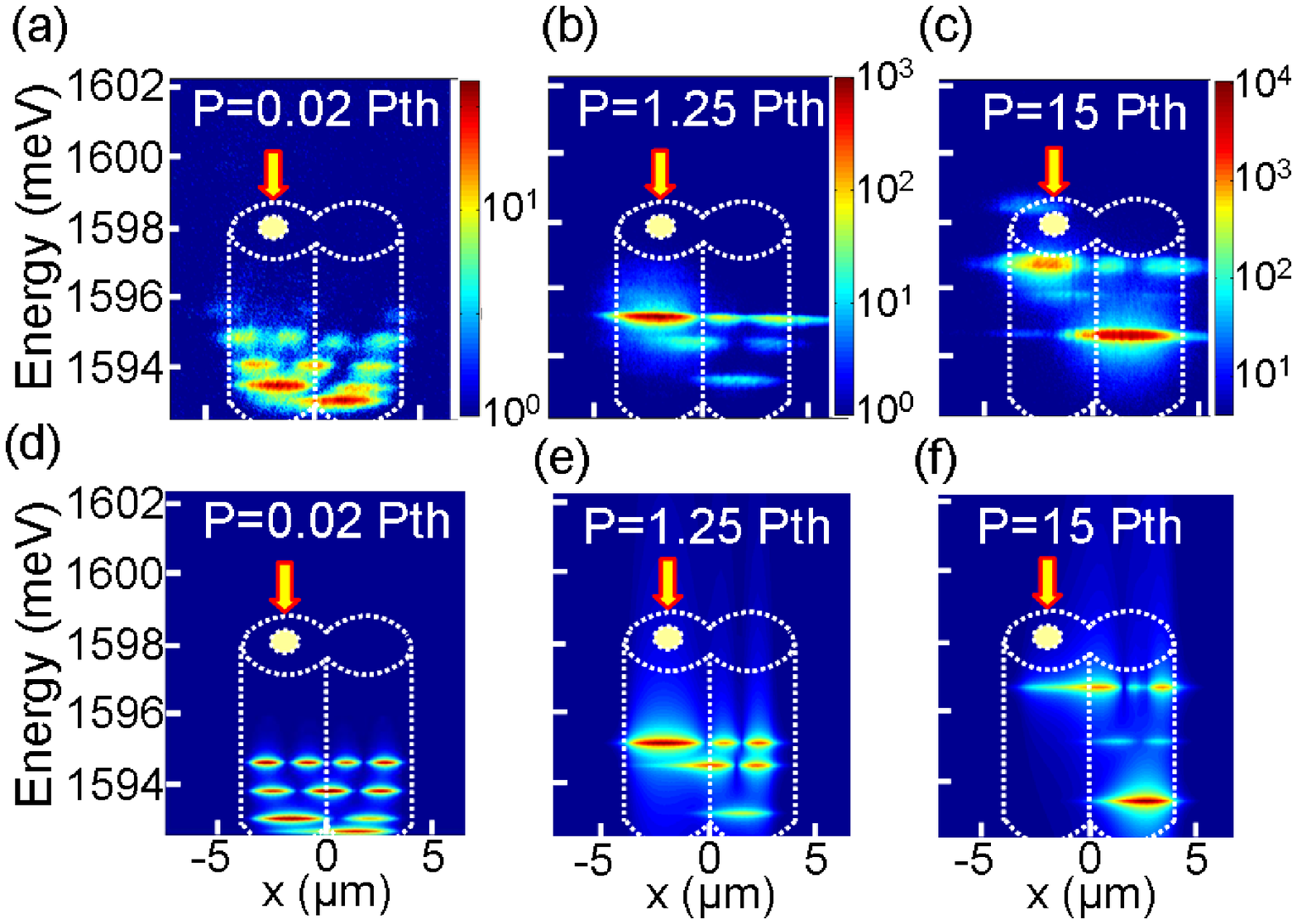}
\caption{(Color online) \textbf{Asymmetric Excitation}:
a)-c)Spectrally resolved emission distribution along the molecule
axis measured for several excitation powers; d)-f) Calculated
emission patterns for several excitation powers. $d\,=\,3.5\,\mu m$
and $d_{CC}\,=\,3.1\,\mu m$. $\delta\,=\,-3\,meV$}\label{Fig4}
\end{figure}

 Different dynamics is observed when the laser spot is positioned
on top of one of the micropillars forming the molecule. Emission
distributions measured under these conditions are reported in
Fig.~\ref{Fig4}.a-c. Close to threshold, condensation occurs in the
pillar that is optically pumped. Interparticle interactions blue
shift the energy of the ground state in the pumped pillar,
decoupling it from the states of the other pillar , and severely
limiting the Josephson transfer of particles from one pillar to the
other. This localisation is very analogous to the original
self-trapping effect, except that here the blue shift is mainly
induced by the interaction between the exciton reservoir and the
condensate and not by interactions within the condensate itself. As
the excitation power is increased, the better relaxation kinetic and
the presence of excited states in the un-pumped pillar destroy the
metastable state\cite{PRLatom} and allow the system reaching its
ground state. Condensation then occurs in the non-pumped micropillar
(Fig.~\ref{Fig4}.c).

 We can reproduce this overall behavior using the self-consistent polariton relaxation model presented
above. As shown in Fig.~\ref{Fig4}.d, at low power the reservoir
potential creates an asymmetry
  in the two bottom states of the molecule, whereas the upper states are not affected. The populations
   of the excitonic states in the reservoir are small and their relative distribution is given by the lattice temperature.
When pumping increases (Fig.~\ref{Fig4}.e), condensation starts at
the  eigenstate possessing the largest overlap integral with the
reservoir. This corresponds to the third polariton state (population
$n_{3}$). The reservoir potential also significantly perturbs the
two bottom states, which become both confined in the right pillar.
 This blocks the direct relaxation from the reservoir towards these states and
 further enforce condensation in the third metastable state
 localized in the pumped pillar.
At high pumping (Fig.~\ref{Fig4}.f) a new effect is observed:
condensation is triggered onto the ground state (of population
$n_{1}$). This does not arise from direct scattering from the
reservoir, since the overlap between the ground state and the
reservoir remains very small. This condensation is the result of
efficient scattering from the intermediate third state toward the
ground state via acoustic phonons.
 This regime occurs when
the phonon assisted scattering rate $W_{XP}n_{3}(1+n_{1})$ from the
intermediate state into the ground state overcomes the ground state
radiative losses $n_{1}\tau_{1}$. This happens when
$n_{3}\approx100$  and further increase of the pumping does not
change $n_{3}$, but only increases $n_{1}$. The second quantized
state remains weakly populated: it has a smaller overlap with the
reservoir
 than the third state and a higher energy than the ground state.
 Overall the complete condensation dynamics can only  be understood considering  the modifications of the relaxation rates
 induced by interactions
 (as shown in the supplementary~\cite{supplementary}
 when constant scattering rates are considered, condensation in the metastable states can not be reproduced).

% This situation is ideal to study the purely
% polariton-polariton interactions, which are usually masked by the
% presence of the dense excitonic reservoir in non-resonant excitation
% experiments.

To conclude, we have demonstrated polariton condensation in photonic
molecules, a fully controlled system in which the coupling constant
between two condensates can be adjusted.  Depending on the precise
location of the excitonic reservoir, strong renormalization of the
polariton states is induced, resulting in strong modification of the
condensation dynamics. In particular condensation can occur in a
metastable localized state , or in the ground state of the system.
These results open the way towards a detailed investigation of
Josephson oscillations~\cite{LagoudakisPRL2011} or, more generally,
 the pumped-dissipative physics of arrays of coupled condensates in an engineered environment.

This work was supported by the C'Nano Ile de France
(Sophiie2), by the ANR (PNANO- 07-005 GEMINI), by the
"Triangle de la Physique" (Picorre), by the FP7 ITN
"Clermont4" (235114) and by the FP7 ITN "Spin-Optronics"(237252).

\end{document}

\title{Supplementary Information on "Polariton Condensation in photonic molecules" }

\author{Marta Galbiati}
\affiliation{CNRS-Laboratoire de Photonique et Nanostructures, Route
de Nozay, 91460 Marcoussis, France}
\author{Lydie Ferrier}
\affiliation{CNRS-Laboratoire de Photonique et Nanostructures, Route
de Nozay, 91460 Marcoussis, France}
\author{Dmitry D. Solnyshkov}
\affiliation{LASMEA, CNRS, Clermont University, University Blaise
Pascal, 24 avenue des Landais, 63177 Aubi\`{e}re cedex, France}
\author{Dimitrii Tanese}
\affiliation{CNRS-Laboratoire de Photonique et Nanostructures, Route
de Nozay, 91460 Marcoussis, France}
\author{Esther Wertz}
\affiliation{CNRS-Laboratoire de Photonique et Nanostructures, Route
de Nozay, 91460 Marcoussis, France}
\author{Alberto Amo}
\affiliation{CNRS-Laboratoire de Photonique et Nanostructures, Route
de Nozay, 91460 Marcoussis, France}
\author{Pascale Senellart}
\affiliation{CNRS-Laboratoire de Photonique et Nanostructures, Route
de Nozay, 91460 Marcoussis, France}
\author{Isabelle Sagnes}
\affiliation{CNRS-Laboratoire de Photonique et Nanostructures, Route
de Nozay, 91460 Marcoussis, France}
\author{Aristide Lema\^{i}tre}
\affiliation{CNRS-Laboratoire de Photonique et Nanostructures, Route
de Nozay, 91460 Marcoussis, France}
\author{Elisabeth Galopin}
\affiliation{CNRS-Laboratoire de Photonique et Nanostructures, Route
de Nozay, 91460 Marcoussis, France}
\author{Guillaume Malpuech}
\affiliation{LASMEA, CNRS, Clermont University, University Blaise
Pascal, 24 avenue des Landais, 63177 Aubi\`{e}re cedex, France}
\author{Jacqueline Bloch}
\affiliation{CNRS-Laboratoire de Photonique et Nanostructures, Route
de Nozay, 91460 Marcoussis,
France}\email[]{jacqueline.bloch@lpn.cnrs.fr}

\date{\today}
\begin{abstract}
We describe in this supplementary information the model we have
developed to simulate polariton relaxation in photonic molecules
under the excitation conditions described in the paper. The results
of these calculations are shown on figure 2 and 4 of the paper.
Finally we show that the overall condensation dynamics is not
reproduced if one does not take into account the interaction induced
changes in the relaxation rates.
\end{abstract}
\maketitle

 We simulate polariton relaxation in
 a self-consistent way, using coupled semi-classical Boltzmann
 equations
 and nonlinear Schr\"{o}dinger equations. At $t=0$ the system is empty and the pumping is switched on. We calculate
 the time evolution of the system until the establishment of a steady state. The nonlinear Schr\"{o}dinger equation written
 below is used to find the spatial shapes and energies $E_{i,t_j}$  of the polariton wave functions $\psi _{i,t_j}(x,y)$.

\begin{equation}
\label{GP}
\begin{gathered}
E_{i,t_j} {\psi _{i,t_j}} =
 - \frac{{{\hbar ^2}}}{{2m}}\Delta {\psi _{i,t_j}} \\
+ \alpha\rho_{R}(x,y) N_R(t_j)
+ \alpha  \left| \sum_{i}\psi _{i,t_{j-1}}e^{-i E_{i,t_{j-1}}t_{j}/\hbar}
\right|^2\psi _{i,t_j}
\end{gathered}
\end{equation}
 m is the polariton effective mass. This equation is solved at time step $t_j$ to find the eigenstates in presence of
the potential created by the exciton reservoir (written as
$\alpha\rho_{R}(x,y) N_R(t_j)$) and by the
occupation of each polariton quantized states (last term of the
equation). $\alpha=6 x E_b (a_B)^2$ is the exciton-exciton interaction constant, where $x$ the exciton fraction of the
considered quantum state (1 in the reservoir, 0.5 in the polariton
states), $E_b$ the exciton binding energy, $a_B$ the exciton Bohr
radius. $\rho_{R}(x,y)$ is the normalized reservoir density
($\int\rho_{R}(x,y)~dxdy=1$), $N_R(t_j)$ the number of excitons in
the reservoir.

The reservoir is assumed to be thermalized at the lattice temperature:
$n_{R}(E,t_j)=n_{R}(0,t_j)\exp\left(-E/k_{B}T\right)$. The total number of particles in the reservoir is
 $N_{R}(t_j)=2\sum_{E}n_{R}(E,t_j)$, where 2 stands for the spin degeneracy of bright heavy-hole excitons.
  Since excitons in the reservoir are confined in a quantum well, $N_{R}(t_j)=n_{R}(0,t_j)Smk_{B}T/2\pi\hbar^2$,
   where $S$ is the surface covered by the reservoir. $\rho_{R}(x,y)$ is assumed to be a Gaussian function
   with a width $w_R$ close to the one of the excitation laser spot.

The time evolution of $n_{i}$, the occupation numbers of the quantized polariton states and of $N_R$, the reservoir population
are calculated using semi-classical Boltzmann equations, taking into account both exciton-exciton and exciton-phonon
interactions, as well as the presence of the thermalized excitonic reservoir.

For the quantized polariton states $i$ the equation reads:
\begin{widetext}
\[\begin{array}{c}
 \frac{{d{n_i}}}{{dt}} =  - \frac{{{n_i}}}{{{\tau _i}}} - {n_i}\sum\limits_k {\left( {{W_{XX}}{I_{ikRR}}{N_R} + {W_{XP}^{ik}}} \right)} ({n_k} + 1)\left( {\begin{array}{*{20}{c}}
   {1,k < i}  \\
   {{e^{ - {E_{ik}}/{k_B}T}},k > i}  \\
\end{array}} \right) \\
  + ({n_i} + 1)\sum\limits_k {\left( {{W_{XX}}{I_{ikRR}}{N_R} + {W_{XP}^{ik}}{I_{ik}}} \right)} {n_k}\left( {\begin{array}{*{20}{c}}
   {{e^{ - {E_{ik}}/{k_B}T}},k < i}  \\
   {1,k > i}  \\
\end{array}} \right) \\
  + \left( {{W_{XX}}{I_{iRRR}}{N_R} + {W_{XP}^{iR}}{I_{iR}}} \right)\left( {{N_R}\left( {{n_i} + 1} \right) - {n_i}\frac{{Smk_{B}T}}{{2\pi {\hbar ^2}}}{e^{ - {E_{iR}}/{k_B}T}}} \right) \\
 \end{array}\]
\end{widetext}

and for the reservoir state $R$ the equation is:
\begin{widetext}
\begin{equation}
\frac{{d{N_R}}}{{dt}} = P - \frac{{{N_R}}}{{{\tau _R}}}- \sum\limits_i {\left( {{W_{XX}}{I_{iRRR}}{N_R} + {W_{XP}^{iR}}{I_{iR}}} \right)\left( {{N_R}\left({{n_i} + 1} \right) - \frac{{{n_i}{S}mk_{B}T}}{{2\pi {\hbar ^2}}}{e^{ - {E_{iR}}/{k_B}T}}} \right)}
\end{equation}
\end{widetext}

$P$ is the non-resonant pumping populating the reservoir, $\tau_R\,=\,400 \,ps$ (resp. $\tau_i\,=\, 30 \,ps$ ) is
the excitonic reservoir (resp. polariton) lifetime.
$W_{XX}$ describes the exciton-exciton
 interaction rates.
The exciton-phonon scattering rate reads:

$$
W_{XP}^{ik,R}=W_{XP}e^{-\frac{(E_{i}-E_{k,R})}{\hbar c\frac{\pi}{a_B}}}
$$

The wave vector conservation rule is replaced by an overlap integral between the spatial extensions of the
initial and final states. The overlap integrals change with time, together with the shapes and the energies
of the states according to equation
(\ref{GP}). The overlap integrals involved are defined as follows:
\begin{equation}
\begin{gathered}
I_{ikRR}=S^3\int\left|\psi_{i}^{*}\psi_{k}\right|^2\rho_{R}^2~dxdy,\\
I_{iRRR}=S^3\int\left|\psi_{i}\right|^2\rho_{R}^{3}~dxdy
\end{gathered}
\end{equation}
The overlap integrals obtained by summation of scattering rates with different phonon wavevectors read:
\begin{equation}
\begin{gathered}
I_{ik}=S\int\left|\psi_{i}^{*}\psi_{k}\right|^2~dxdy,\\
I_{iR}=S\int\left|\psi_{i}\right|^2\rho_{R}~dxdy
\end{gathered}
\end{equation}
These overlap integrals are dimensionless.

Finally, the scattering rates $W_{XX}$ and $W_{XP}$ describing exciton-exciton and exciton-phonon interaction were used as the fitting parameters of the model, together with the width of the gaussian spatial distribution of the reservoir which we authorize to slightly vary above the width given by the pumping laser spot size.
We used $W_{XX}=2~10^3~s^{-1}$, $W_{XP}=10^9~s^{-1}$, $w_r=3 \mu m$ .

In order to understand the terms in the Boltzmann equation, let us consider for example the case of the scattering from the reservoir towards a quantized state $i$ and backward assisted by exciton-phonon
and exciton-exciton interaction. We neglect the final state stimulation within the reservoir which means that we assume a weak occupation ($n_R(E)\ll 1$) of these states.
 In all cases, the energy difference between the state $i$ and the bottom of the exciton reservoir is $E_0$.

 \begin{figure}[t]
\includegraphics[width=0.2\columnwidth]{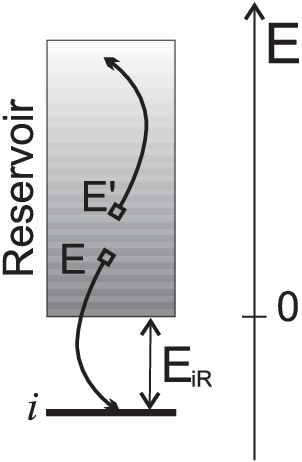}
\caption{Schematic representation of the exciton-exciton interaction in the reservoir leading to the scattering towards the quantized polariton states. The reservoir has a 2D (constant) density of states with a thermal distribution function.}\label{Fig1c}
\end{figure}

  Let us first consider the most complicated case of the forward scattering assisted
  by  exciton-exciton interaction (sketched in figure S-1). In order to calculate the corresponding
scattering rate, we have to sum over all possible scattering events. The corresponding scattering rates reads:

\begin{widetext}
\begin{equation}
\begin{gathered}
W_{XX}I_{iRRR}\left(n_i + 1\right)\sum\limits_{E = 0}^\infty  n_R\left(E\right)\sum\limits_{E'}^\infty  n_R\left( E' \right)\left( 1 + n_R\left( E' + E + E_{iR} \right) \right) \approx W_{XX}I_{iRRR}\left( n_i + 1 \right)N_R^2
\end{gathered}
\end{equation}
\end{widetext}
The rate is proportional to $N_R^2$ and is similar to a scattering rate written between individual states and based on particle-particle interaction.
The backward process of one particle going from the quantized states to the reservoir  reads:

\begin{widetext}
\begin{equation}
\begin{gathered}
\\
 W_{XX}I_{iRRR}n_i\sum \limits_{E = 0}^\infty  (1 + n_R\left( E \right))\mathop \sum \limits_{E' = 0}^\infty  n_R(E' + E + E_{iR})(1 + n_R(E')) \hfill \\
   \approx W_{XX}I_{iRRR}n_i\mathop \sum \limits_{E = 0}^\infty  \left( 1 + n_R\left( E \right) \right) e^{ - \frac{E + E_{iR}}{kT}}\mathop \sum \limits_{E' = 0}^\infty  n_R(E') \hfill \\
   \approx W_{XX}I_{iRRR}n_i N_R e^{ - \frac{E_{iR}}{kT}}\mathop \sum \limits_{E = 0}^\infty  e^{ - \frac{E}{kT}} = W_{XX}I_{iRRR} n_i\frac{N_R^2}{n_R(0)} e^{ - \frac{E_{iR}}{kT}} = W_{XX}I_{iRRR}\frac{S mkT}{2\pi \hbar^2}n_i N_R e^{ - \frac{E_{iR}}{kT}} \hfill \\
\end{gathered}
\end{equation}
\end{widetext}
The rate is similar to the one of the forward process except by the term : $e^{ - \frac{E_{iR}}{k_{B}T}}/n_R(0)$.
The relaxation of particles toward a low energy state is therefore favored only for a minimum energy difference $E_0>k_{B}T~\ln(n_R(0))$.

For the phonon-assisted scattering between the reservoir $R$ and a quantized state $i$ the expressions are quite similar. The scattering rates from the reservoir towards the state $i$ reads:
\begin{widetext}
\begin{equation}
\begin{gathered}
{W_{XP}}{I_{iR}}\left( {{n_i} + 1} \right)\sum\limits_E {n_R}\left( E \right)\left( 1 + {n_{ph}}\left( {E + {E_{iR}}} \right) \right)  \approx {W_{XP}}{I_{iR}}\left( {{n_i} + 1} \right){N_R}
\end{gathered}
\end{equation}
\end{widetext}

The scattering rates from the state $i$ toward the reservoir reads:
\begin{widetext}
\begin{equation}
\begin{gathered}
{W_{XP}}{I_{iR}}{n_i}\sum\limits_{E = 0} {{n_{ph}}\left( E \right)\left( {1 + {n_R}\left( E \right)} \right)}  \approx {W_{XP}}{I_{iR}}{n_i}{e^{ - {E_{iR}}/{k_B}T}}\frac{Sk_{B}T}{2\pi\hbar^2}
\end{gathered}
\end{equation}
\end{widetext}

One can note that the expressions for the scattering processes involving exciton reservoir
 and phonons are quite similar. This is explained by the fact that excitonic reservoir is
 assumed to be thermalized at the lattice temperature.

Despite the complexity of the equations used in our simulation, we
would like to underline that the number of fitting parameter is
really limited.

\begin{figure}[t]
\includegraphics[width=0.8\columnwidth]{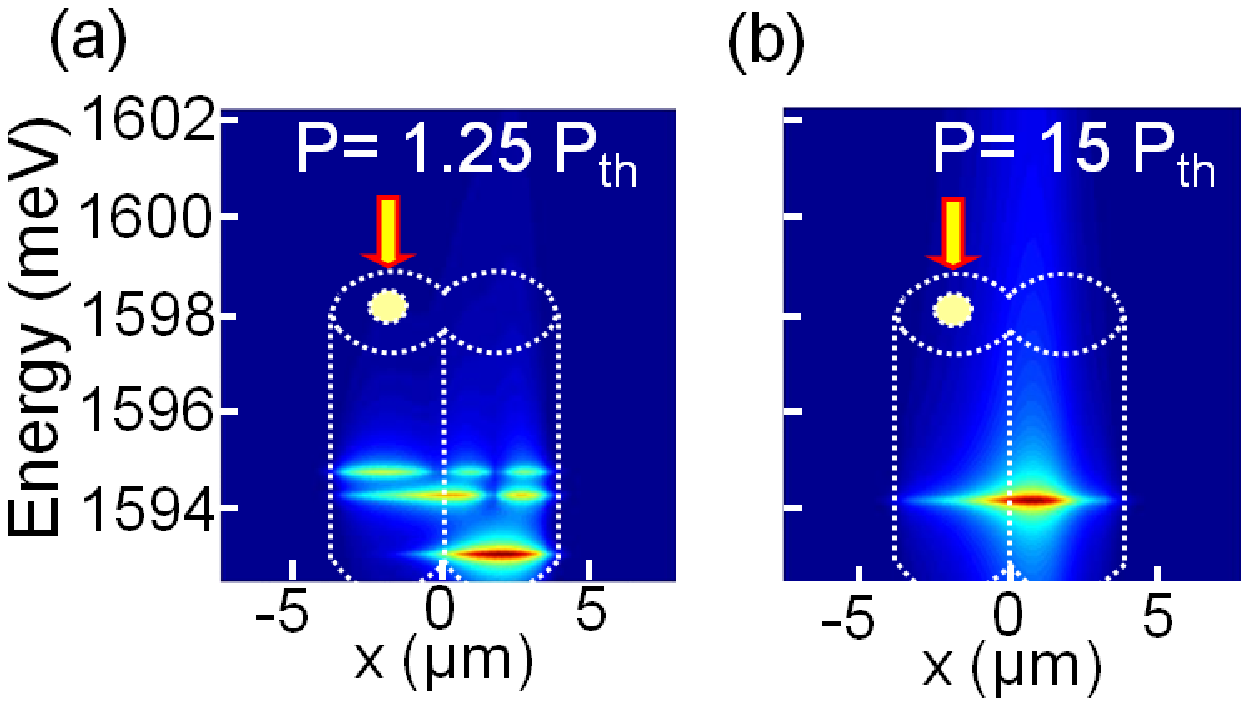}
\caption{Calculated emission pattern considering a reservoir
potential proportional to the excitation power in the left
micropillar but without taking into account the changes in the
scattering rates.}\label{FigB}
\end{figure}

Figure\ref{FigB} presents calculations corresponding to the
experiments described in fig.4 of the paper. The reservoir is
injected in the left pillar. The difference with the calculations
presented in fig.4 is that here the scattering rates are taken
constant, equal to that obtained at low excitation density. Thus we
neglect the effect of the polariton wavefunction renormalization on
the scattering rates. Clearly the main features of the experiments
are not well described, in particular condensation in the metastable
state at intermediate excitation powers is not reproduced.

 The crucial ingredient which allows catching the basic physics in our experiments is
to take into account the overlap integrals between initial and final
states in the scattering rates. As shown in the paper, these wave
functions are strongly distorted by interactions, and this
dramatically modifies the way polariton relaxation occurs in the
system.